\documentclass[a4paper, conference]{IEEEtran}
\IEEEoverridecommandlockouts
\usepackage[utf8]{inputenc}
\usepackage{amsmath,amsfonts}
\usepackage{algorithm}
\usepackage{algpseudocode}
\usepackage{booktabs} 
\usepackage{array}    
\usepackage{tabularx}
\usepackage[caption=false,font=normalsize,labelfont=sf,textfont=sf]{subfig}
\usepackage{textcomp}
\usepackage{stfloats}
\usepackage{url}
\usepackage{verbatim}
\usepackage{graphicx}
\usepackage{cite}
\usepackage{graphicx}
\usepackage{tikz}
\usepackage{pgfplots}
\usepackage{psfrag}
\usepackage[justification=centering]{caption}
\usetikzlibrary{arrows,snakes,shapes,shapes.geometric,fit,positioning,arrows,decorations.markings,calc}
\usepackage{mathtools}
\usepackage{amsbsy}
\usepackage{amsthm}
\usepackage{fixmath}
\usepackage{amssymb}
\usepackage{mathrsfs}
\usepackage{fancyhdr}
\usepackage{xcolor}
\usepackage{hyperref}
\usetikzlibrary{spy}

\newcommand{\HH}{\mathrm{H}}

\newcommand{\tdl}{\mathrm{dl}}
\newcommand{\bs}[1]{\boldsymbol{#1}}

\newcommand{\tr}{\text{tr}}

\newtheorem{proposition}{Proposition}[section]
\definecolor{lime}{HTML}{A6CE39}

\DeclareRobustCommand{\orcidicon}{%
	\begin{tikzpicture}
	\draw[lime, fill=lime] (0,0) 
	circle [radius=0.16] 
	node[white] {{\fontfamily{qag}\selectfont \tiny ID}};
	\draw[white, fill=white] (-0.0625,0.095) 
	circle [radius=0.007];
	\end{tikzpicture}
	\hspace{-2mm}
}

\foreach \x in {A, ..., Z}{%
	\expandafter\xdef\csname orcid\x\endcsname{\noexpand\href{https://orcid.org/\csname orcidauthor\x\endcsname}{\noexpand\orcidicon}}
}

\makeatletter
\def\endthebibliography{%
	\def\@noitemerr{\@latex@warning{Empty `thebibliography' environment}}%
	\endlist
}
\makeatother

\begin{document}
	
	\title{Robust Precoding for FDD MISO Systems via Minorization Maximization
 \thanks{This manuscript has been accepted for publication at VTC Fall 2024}
	}
	
	\author{\IEEEauthorblockN{Donia~Ben~Amor\orcidA{}, Michael~Joham\orcidB{}, Wolfgang~Utschick\orcidC{}}
		\IEEEauthorblockA{ 	\textit{Chair of Methods of Signal Processing} \\
			\textit{School of Computation, Information and Technology, Technical University of Munich} \\
			Munich, Germany \\
			Email: \{donia.ben-amor, joham, utschick\}@tum.de}
	}
	\maketitle

\begin{abstract}
In this work, we propose an approach to robust precoder design based on a minorization maximization technique that optimizes a surrogate function of the achievable spectral efficiency. The presented method accounts for channel estimation errors during the optimization process and is, hence, robust in the case of imperfect channel state information (CSI). Additionally, the design method is adapted such that the need for a line search to satisfy the power constraint is eliminated, that significantly accelerates the precoder computation. Simulation results demonstrate that the proposed robust precoding method is competitive with weighted minimum mean square error (WMMSE) precoding, in particular, under imperfect CSI scenarios.
\end{abstract}
	
	
\section{Introduction}
Precoding techniques for frequency-division-duplex (FDD) multi-user (MU) multiple-input single-output (MISO) systems have been the subject of intense research due to their potential to enhance spectral efficiency and system performance. In FDD systems, the uplink and downlink transmissions occur on separate frequency bands, presenting unique challenges for channel state information (CSI) acquisition and precoder design \cite[Subsection 5.4.6]{Tse}.

A key challenge in FDD systems is obtaining accurate CSI at the transmitter, as the channel reciprocity that exists in time division duplex systems cannot be exploited. This has led researchers to explore various approaches for CSI feedback and precoding strategies tailored to FDD scenarios.

Traditional precoding schemes like zero-forcing (ZF) \cite{Wiesel} and minimum-mean-square-error (MMSE) precoding \cite{MMSEprecoder} have been widely studied for MU-MISO systems. However, these techniques can face limitations in certain scenarios, particularly when the number of training pilots is very limited---a scenario we consider in this article.

Recent works have focused on leveraging optimization techniques to maximize the sum rate for linear precoding in MU-MISO systems using various methods such as the iterative weighted minimum mean square error (IWMMSE) algorithm \cite{IWMMSE_Luo}, the fractional programming framework \cite{FracProg}, and the minorization maximization (MM) approach \cite{MM}.

The MM algorithm, proposed in \cite{MM}, has shown promising results in maximizing the weighted sum rate with linear precoding for both MISO and MIMO systems in the perfect CSI case. This algorithm exploits the flexibility of the MM framework to construct an effective surrogate function, leading to an efficient optimization procedure.

Building upon this foundation, our work extends the MM approach to address the more challenging scenario of imperfect CSI in FDD MU-MISO systems. We propose a robust precoding approach that incorporates channel estimation errors into the optimization process, making it particularly suitable for the practical case of imperfect CSI. To this end, we construct a novel surrogate function for the training-based achievable sum rate expression. The latter is based on an error model for the channel accounting for the errors resulting from linear MMSE (LMMSE) channel estimation \cite{Hasssibi_Hochwald,VTC24}. 

A key contribution of our work is the elimination of the line search procedure typically required to satisfy the power constraint. The idea is inspired by the scaling approach first proposed in \cite{JoUtNo05} and later used in \cite{RethinkWMMSE}. In a nutshell, the optimization problem with respect to the precoders is reformulated as an unconstrained optimization, which allows a closed-form solution for the precoder update within the iterative procedure. This leads to a significant reduction of the computational complexity of the algorithm, addressing a limitation noted in the MM approach \cite[beginning of Section VI]{MM}.

Our method not only builds upon the strengths of the MM framework but also addresses its limitations, particularly in scenarios with imperfect CSI. By accounting for channel estimation errors and eliminating the need for a line search, our approach offers a more robust and computationally efficient solution for precoding in FDD MU-MISO systems.

We show through numerical results the competitiveness of the MM method based on the sum rate lower bound (MM-LB) to the augmented weighted average MSE (AWAMSE) approach proposed in our recent work \cite{VTC24}. The advantage of the closed-form precoder update is also demonstrated as compared to the approach based on line search via bisection (MMbisec-LB). 

Furthermore, we implemented an adapted version of the "MMplus" method proposed in \cite[Section VI]{MM} which circumvents the line search by considering a lower bound on the surrogate function. The so-obtained optimization problem is drastically simplified and can be solved in closed form. Simulation results show, however, that this approach requires much more iterations to converge and is, in most cases, not as computationally efficient as the proposed MM-LB algorithm.


\section{System Model}\label{sec:SysModel}
Consider the downlink of a MU-MISO FDD system, where the base station (BS) is equipped with $M\gg 1$ antennas and serves $K$ single-antenna users. Unlike time division duplex systems, there is no reciprocity between the uplink and downlink channels in the FDD setup. Hence, the BS needs to explicitly probe the downlink channel by sending $T_\tdl$ pilots to the users. The $T_\tdl$ channel observations received at the users' sides are collected and then fed back to the BS via an analog feedback channel. At the BS, we obtain 
\begin{equation}
    \bs{y}_k = \bs{\Phi}^\HH \bs{h}_k + \bs{n}_k.
\end{equation}
Here, $\bs{h}_k\sim \mathcal{N}_\mathbb{C}(\bs{0}, \bs{C}_k)$ denotes the channel between the BS and the $k$-th user. The pilot matrix with unit-norm columns is denoted as $\bs{\Phi} \in \mathbb{C}^{M\times T_\tdl}$.  The additive white Gaussian noise is $\bs{n}_k\sim \mathcal{N}_\mathbb{C}(\bs{0},\frac{1}{P_\tdl}\bs{I})$ resulting from the noisy downlink training and feedback, where $P_\tdl$ denotes the downlink transmit power. 

Note that we are particularly interested in the case where $T_\tdl<M$ in order to reduce the training overhead, i.e., the number of pilots is smaller than the number of antennas. Such an assumption leads to the difficult situation of incomplete channel knowledge at the BS.

Based on the observation $\bs{y}_k$, the BS can compute a channel estimate $\hat{\bs{h}}_k$  and the channel vector can therefore be modeled as follows
\begin{equation}
    \bs{h}_k= \hat{\bs{h}}_k + \Tilde{\bs{h}}_k \label{eq:hk}
\end{equation}
where $\Tilde{\bs{h}}_k$ stands for the zero-mean estimation error whose covariance matrix is denoted by $\bs{C}_{\text{err},k}$. \\
Assuming that LMMSE channel estimation is performed, $\hat{\bs{h}}_k$ can be written as
\begin{equation}
\begin{aligned}
    \hat{\bs{h}}_k &= \mathbb{E}[\bs{h}_k|\bs{y}_k] =\mathbb{E}[\bs{h}_k \bs{y}_k^\HH] \mathbb{E}[\bs{y}_k \bs{y}_k^\HH]^{-1} \bs{y}_k\\
    &=\bs{C}_k \bs{\Phi} \left(\bs{\Phi}^\HH \bs{C}_k \bs{\Phi} +\frac{1}{P_\tdl} \bs{I}\right)^{-1} \bs{y}_k.
\end{aligned}
\end{equation}
Due to the orthogonality principle \cite[Section 8.1]{Scharf1989}, the LMMSE channel estimate $\hat{\bs{h}}_k$ and the corresponding estimation error vector $\Tilde{\bs{h}}_k$ are uncorrelated. Furthermore, the error covariance matrix reads as
\begin{equation}
\begin{aligned}
     \bs{C}_{\text{err},k} &= \mathbb{E}[\Tilde{\bs{h}}_k \Tilde{\bs{h}}_k^\HH] \\
     &=\bs{C}_k - \bs{C}_k \bs{\Phi}  \left(\bs{\Phi}^\HH \bs{C}_k \bs{\Phi} +\frac{1}{P_\tdl} \bs{I}\right)^{-1} \bs{\Phi}^\HH \bs{C}_k.
\end{aligned}
\end{equation}

During the data transmission phase, user $k$ receives the signal $r_k$ given by
\begin{equation} \label{eq:rk}
    r_k = \bs{h}_k^\HH \bs{p}_k s_k + \sum_{j\neq k} \bs{h}_k^\HH \bs{p}_j s_j + v_k
\end{equation}
where $s_i \sim \mathcal{N}_\mathbb{C}(0,1)$ denotes the data symbol intended for user $i$ and $v_k \sim \mathcal{N}_\mathbb{C}(0,1)$ stands for the AWGN at the $k$-th user antenna. The $\bs{p}_k$ is the precoding vector for user $k$ that shall be optimized such that the spectral efficiency of the system is maximized under a transmit power constraint.

\section{Optimization Problem}\label{sec:Fig-of-Merit}
Before presenting the optimization of the precoding vectors, we firstly formulate our figure of merit and then present our solution approach to the underlying optimization problem. 

Note that in the case of a perfect CSI, the achievable spectral efficiency of user $k$ is computed as follows
\begin{equation}\label{eq:RkPerfectCSI}
    R_k = \mathbb{E}[\log_2(1+\gamma_k)]
\end{equation}
where the expectation is with respect to the channel realizations. The signal-to-interference-and-noise ratio (SINR) $\gamma_k$ is given by
\begin{equation}
    \gamma_k = \frac{| \bs{h}_k^\HH \bs{p}_k|^2}{\sum_{j\neq k} |\bs{h}_k^\HH \bs{p}_j|^2 + 1}.
\end{equation}
However, since we lack complete channel knowledge in the considered setup, relying on \eqref{eq:RkPerfectCSI} to design the precoding vectors can lead to poor performance. This is because the channel estimates are treated as if they were the true channels and no estimation error is accounted for during the design.

Hence, we formulate a lower bound on the achievable spectral efficiency taking into account the channel estimation errors (cf. \cite{Hasssibi_Hochwald,VTC24}) and use it as our figure of merit. 

Based on the channel model in \eqref{eq:hk}, the received signal \eqref{eq:rk} can be decomposed as follows
\begin{equation} 
    r_k = \hat{\bs{h}}_k^\HH \bs{p}_k s_k + \Tilde{\bs{h}}_k^\HH \bs{p}_k s_k + \sum_{j\neq k} \bs{h}_k^\HH \bs{p}_j s_j + v_k.
\end{equation}
Assuming LMMSE channel estimation is performed and due to the Gaussianity of $\bs{h}_k$ and $\bs{n}_k$, the achievable spectral efficiency of user $k$ is lower bounded as follows (cf. \cite{Hasssibi_Hochwald,VTC24})
\begin{equation}\label{eq:LB}
    C_k\geq \Tilde{R}_k=\mathbb{E}[\log_2(1+\bar{\gamma}_k)]
\end{equation}
where the expectation is with respect to the channel realizations. The SINR $\bar{\gamma}_k$ introduced in \eqref{eq:LB} is defined as
\begin{equation}
    \bar{\gamma}_k = \frac{|\hat{\bs{h}}_k^\HH \bs{p}_k|^2}{\sum_{j\neq k} |\hat{\bs{h}}_k^\HH \bs{p}_j|^2 + \sum_{j=1}^K \bs{p}_j^\HH \bs{C}_{\text{err},k}\bs{p}_j + 1}
\end{equation}
where the independence between the channel estimate and the estimation error was used (see \cite[Section 8.1]{Scharf1989}).

The optimization problem can now be formulated as
\begin{equation}\label{eq:OP_LB}
  \underset{\bs{P}}{\max} \: \sum_{k=1}^K  \log_2(1+\bar{\gamma}_k) \quad \text{s.t.} \quad \|\bs{P}\|_\mathrm{F}^2 \leq P_\tdl
\end{equation}
where we introduced the precoding matrix $\bs{P}=[\bs{p}_1, \dots, \bs{p}_K]$.\\
Note that \eqref{eq:OP_LB} is non-convex due to the fraction in the log function and the dependence of the individual users' rates on the precoding vectors of all users. 

\section{Precoder Design via MM}\label{sec:MM}
Inspired by the approach in \cite{MM} developed perfect CSI, we propose to apply a minorization maximization algorithm, by first defining a surrogate function for the objective in \eqref{eq:OP_LB} that serves as a lower bound to the spectral efficiency and then maximizing the surrogate function in a second step. 

To this end, we state the following proposition that will help us provide a lower bound on the achievable rate 
\begin{proposition}\cite[Section III.B]{MM-RIS}\\
\label{prop:lowerbound}
    For some $x\in \mathbb{C}$ and $z>0$, it holds that\\
    \begin{equation}\label{eq:PropIneq}
    \begin{aligned}
         \log_2(1+\frac{|x|^2}{z})\geq  &\log_2(1+\frac{|\bar{x}|^2}{\bar{z}})
-\frac{|\bar{x}|^2}{\bar{z}} + 2\Re\{\frac{\bar{x}^*}{\bar{z}}x\} \\- &\frac{|\bar{x}|^2}{\bar{z}(\bar{z}+|\bar{x}|^2)} (z + |x|^2)
    \end{aligned}
    \end{equation}
    with equality for $(x,z)=(\bar{x},\bar{z})$.
\end{proposition}
Unlike \cite{MM} and in order to derive a closed-form update equation for the precoder, we perform the following substitutions
\begin{equation}
        x=\beta^{-\frac{1}{2}} \hat{\bs{h}}_k^\HH \bs{p}_k,\: z=\beta^{-1}\Big(\sum_{j \neq k}|\hat{\bs{h}}_k^\HH \bs{p}_j|^2 + \sum_{j=1}^K \bs{p}_j^\HH \bs{C}_{\text{err},k} \bs{p}_j+1\Big) \label{eq:Subst}
\end{equation}
\begin{equation}
    \bar{x}=\hat{\bs{h}}_k^\HH \bar{\bs{p}}_k, \: \bar{z}=\sum_{j \neq k}|\hat{\bs{h}}_k^\HH \bar{\bs{p}}_j|^2 + \sum_{j=1}^K \bar{\bs{p}}_j^\HH \bs{C}_{\text{err},k} \bar{\bs{p}}_j+1
\end{equation}
where we additionally introduced the scaling factor $\beta>0$ which will be determined in the sequel.

Note that for any $\beta>0$, the substitutions in \eqref{eq:Subst} lead to the lower bound on the achievable rate of user $k$ defined in \eqref{eq:LB}, i.e., $\log_2(1+\frac{|x|^2}{z})=\bar{R}_k$.

The surrogate function for the achievable rate of user $k$ at $\bar{\bs{P}}$ is therefore given by
    \begin{align}
        f_k(\bs{P},\bar{\bs{P}},\beta) &= \bar{R}_k(\bar{\bs{P}}) - \bar{\gamma}_k(\bar{\bs{P}}) + 2 \beta^{-\frac{1}{2}} \Re\{b_k\hat{\bs{h}}_k^\HH \bs{p}_k \} \label{eq:SFk} \\
        &-  \beta^{-1} a_k \Big( \sum_{j=1}^K |\hat{\bs{h}}_k^\HH \bs{p}_j|^2  + \sum_{j=1}^K \bs{p}_j^\HH \bs{C}_{\text{err},k} \bs{p}_j+1\Big) \nonumber
    \end{align}
where we introduced the coefficients $a_k$ and $b_k$ 
\begin{align}
    a_k &= \frac{\bar{\gamma}_k(\bar{\bs{P}})}{ \sum_{j=1}^K |\hat{\bs{h}}_k^\HH \bs{p}_j|^2 + \sum_{j=1}^K \bs{p}_j^\HH \bs{C}_{\text{err},k} \bs{p}_j+1} \label{eq:ak}\\
    b_k &= \frac{\bar{\gamma}_k(\bar{\bs{P}})}{ \hat{\bs{h}}_k^\HH \bar{\bs{p}}_j }.\label{eq:bk}
\end{align}
A surrogate function, which by Proposition~\ref{prop:lowerbound} is a lower bound, for the sum rate can be obtained as the sum of the $K$ terms defined in \eqref{eq:SFk} and can be written as
\begin{align}
     f(\bs{P},\bar{\bs{P}},\beta)&=\sum_{k=1}^K   f_k(\bs{P},\bar{\bs{P}},\beta)\label{eq:mySurr} \\
     &=\Big(\sum_{k=1}^K \bar{R}_k(\bar{\bs{P}}) - \bar{\gamma}_k(\bar{\bs{P}}) \Big)
    +2 \beta^{-\frac{1}{2}} \Re\{\bs{B} \bs{\hat{H}}^\HH \bs{P}\} \nonumber \\
    &- \beta^{-1} \Big( \tr(\bs{A} \bs{\hat{H}}^\HH \bs{P} \bs{P}^\HH \bs{\hat{H}}) +\tr(\bs{P} \bs{P}^\HH \bs{Z})+\tr(\bs{A})\Big). \nonumber
\end{align}
Here, we defined the diagonal matrices $\bs{A}=\text{diag}(a_1,\dots, a_K)$ and $\bs{B}=\text{diag}(b_1,\dots, b_K)$. Additionally, we introduced the estimated channel matrix $\bs{\hat{H}}=[\bs{\hat{h}}_1,\dots, \bs{\hat{h}}_K]$ and 
\begin{equation}
  \bs{Z}=\sum_{i=1}^K a_i \bs{C}_{\text{err},i}.
  \label{eq:sumoferrorcov}
\end{equation}

Hence, we can state the following maximization problem of the surrogate function
\begin{equation}\label{eq:OP_surrogate}
  \underset{\bs{P}, \beta}{\max} \: f(\bs{P},\bar{\bs{P}},\beta) \quad \text{s.t.} \quad \|\bs{P}\|_\mathrm{F}^2 \leq P_\tdl.
\end{equation}
By introducing the Lagrangian multiplier $\lambda\geq 0$ associated with the power constraint, we can formulate the Lagrangian function corresponding to \eqref{eq:OP_surrogate}
\begin{equation}
    \mathcal{L}(\bs{P},\beta, \lambda)=f(\bs{P},\bar{\bs{P}},\beta) - \lambda (\|\bs{P}\|_\mathrm{F}^2 - P_\tdl).
\end{equation}
The first optimality condition yields the following expression for the precoding matrix
\begin{equation}\label{eq:P_delta}
    \bs{P}(\beta)=\beta^{\frac{1}{2}} \left( \bs{Z} + \hat{\bs{H}}\bs{A}\hat{\bs{H}}^\HH + \delta \bs{I} \right)^{-1} \hat{\bs{H}}\bs{B}^*
\end{equation} 
where we defined the scaled Lagrangian multiplier $\delta = \lambda \beta$.

The optimal scaling factor $\beta$ can be inferred from the power constraint (is an equality in the optimum: $\|\bs{P}(\beta)\|_\mathrm{F}^2=P_\tdl$), and can be expressed as a function of $\delta$
\begin{equation}\label{eq:beta}
    \beta(\delta)= \frac{P_\tdl}{\tr\left(\bs{X}(\delta)^{-2}\hat{\bs{H}}\bs{B}^*\bs{B}\hat{\bs{H}}^\HH\right)}
\end{equation}
with $\bs{X}(\delta)=\bs{Z} + \hat{\bs{H}}\bs{A}\hat{\bs{H}}^\HH + \delta \bs{I}$ [see also \eqref{eq:sumoferrorcov}]

Given the solution of the precoder in \eqref{eq:P_delta} and the scaling factor in \eqref{eq:beta} as a function of $\delta$, we now obtain the following unconstrained optimization problem with respect to $\delta$
\begin{equation}\label{eq:OP_delta}
    \begin{aligned}
        \underset{\delta}{\max} \: &2\Re\{\tr(\bs{B}\hat{\bs{H}}^\HH \bs{X}(\delta)^{-1}\hat{\bs{H}}\bs{B}^*)\}\\
        -&\tr(\bs{A}\hat{\bs{H}}^\HH \bs{X}(\delta)^{-1} \hat{\bs{H}}\bs{B}^* \bs{B} \hat{\bs{H}}^\HH \bs{X}(\delta)^{-1} \hat{\bs{H}})\\
        -&\tr(\bs{X}(\delta)^{-1}\hat{\bs{H}}\bs{B}^* \bs{B} \hat{\bs{H}}^\HH \bs{Z})\\
        -&\frac{1}{P_\tdl}\tr(\bs{X}(\delta)^{-2}\hat{\bs{H}}\bs{B}^* \bs{B} \hat{\bs{H}}^\HH) \,\tr(\bs{A}).
    \end{aligned}
\end{equation}
The solution to \eqref{eq:OP_delta} is obtained by setting the first derivative of the objective to zero and reads as
\begin{equation}
    \delta = \frac{\tr(\bs{A})}{P_\tdl}.
\end{equation}
The closed-form expression for the unscaled precoder is therefore given by
\begin{equation}\label{eq:P_unscaled}
    \bs{P}^\text{unscaled}= \left( \bs{Z} + \hat{\bs{H}}\bs{A}\hat{\bs{H}}^\HH +  \frac{\tr(\bs{A})}{P_\tdl} \bs{I} \right)^{-1} \hat{\bs{H}}\bs{B}^*.
\end{equation}
The overall procedure is summarized in Algorithm~\ref{alg:1}. Note that by the comment below \eqref{eq:Subst} it is sufficient to work with $\bs{P}^\text{unscaled}$ in the iteration since the scaling by $\beta$ in the surrogate function does not effect the achievable rate.
\begin{algorithm}
\caption{MM-LB Algorithm}
\label{alg:1}
\begin{algorithmic}[1] 
\State  Initialize the precoding matrix $\bs{P}$ 
\State Compute the coefficients $a_k$ and $b_k$ according to \eqref{eq:ak}
 and \eqref{eq:bk}, respectively 
\State Calculate the precoding matrix $\bs{P}^\text{unscaled}$ based on \eqref{eq:P_unscaled}
\State Repeat steps 2-3 until convergence of the sum rate lower bound
\State Scale the precoding matrix with $\beta^{1/2}$ [see \eqref{eq:beta}] to satisfy the power constraint
\end{algorithmic}
\end{algorithm}

\subsection{Adaptation of the MMplus Approach}
As in \cite[Section VI]{MM}, we present a variant of the MM approach which exhibits a closed-form update of the precoder. To this end, an additional surrogate function for the one given in \eqref{eq:SFk} is constructed. Specifically, the sum of the quadratic terms in \eqref{eq:SFk}, i.e., 
$\sum_{k=1}^K (-a_k \sum_{j=1}^K |\hat{\bs{h}}_k^\HH \bs{p}_j|^2 -a_k \sum_{j=1}^K \bs{p}_j^\HH   \bs{C}_{\text{err},k}\bs{p}_j )$ has now to be lower bounded.

Firstly, we rewrite this term as follows
\begin{equation}\label{eq:Quad2}
    -\sum_{k=1}^K  \bs{p}_k^\HH \underbrace{\Big(\sum_{j=1}^K  a_j\hat{\bs{h}}_j\hat{\bs{h}}_j^\HH \Big)}_{\bs{L}_1} \bs{p}_k -  \sum_{k=1}^K \bs{p}_k^\HH \underbrace{\Big(\sum_{j=1}^K  a_j\bs{C}_{\text{err},j}\Big)}_{\bs{L}_2}  \bs{p}_k 
\end{equation}
Let $\eta_1\geq \lambda_\text{max}(\bs{L}_1)$ and $\eta_2\geq \lambda_\text{max}(\bs{L}_2)$, then we can use \cite[Proposition 16]{MM} with $\bs{M}_1=\eta_1 \bs{I}$ and $\bs{M}_2=\eta_2\bs{I}$ to provide a lower bound for the two quadratic terms in \eqref{eq:Quad2}. The scalars $\eta_1$ and $\eta_2$ can be found with $\eta_1= \sum_{j=1}^K  a_j \|\hat{\bs{h}}_j\|^2, \:
    \eta_2= \sum_{j=1}^K  a_j \|\bs{C}_{\text{err},j}\|_\mathrm{F}$. With $\eta=\eta_1+\eta_2 $ we obtain the following 
surrogate function to the surrogate function in \eqref{eq:mySurr}
\begin{equation}\label{eq:Surrogate}
 \begin{aligned}
   & f'(\bs{P}, \bar{\bs{P}}) =-\sum_{k=1}^K \eta\bs{p}_k^\HH\bs{p}_k +\sum_{k=1}^K  2\Re\{ b_k \bs{h}_k^\HH \bs{p}_k\}\\&-\sum_{k=1}^K  2\Re\{ \bar{\bs{p}}_k^\HH (\sum_{j=1}^K  a_j(\hat{\bs{h}}_j\hat{\bs{h}}_j^\HH +\bs{C}_{\text{err},j})\eta\bs{I})\bs{p}_k\}  \\
&-\sum_{k=1}^K\bar{\bs{p}}_k^\HH \Big(\eta\bs{I}-\sum_{j=1}^K  a_j(\hat{\bs{h}}_j\hat{\bs{h}}_j^\HH +\bs{C}_{\text{err},j})\Big)\bar{\bs{p}}_k\\
&+ \sum_{k=1}^K \left(  \bar{R}_k(\{\bar{\bs{p}}_i\}) - \bar{\gamma}_k(\{\bar{\bs{p}}_i\}) -a_k \right).
\end{aligned}   
\end{equation}
By defining $\bs{q}_k$ as
\begin{equation}
    \bs{q}_k = \eta^{-1} \Bigg( b_k^* \bs{h}_k - \Big(\sum_{j=1}^K  a_j (\hat{\bs{h}}_j \hat{\bs{h}}_j^\HH + \bs{C}_{\text{err},j}) - \eta\bs{I}\Big)\bar{\bs{p}}_k\Bigg)
\end{equation}
we end up with the following optimization problem
\begin{equation}
    \underset{\{\bs{p}_i\}}{\min} \: \sum_{k=1}^K \eta \|\bs{p}_k -\bs{q}_k \|^2
\end{equation}
whose solution is given in closed form as follows
\begin{equation}
    \bs{p}_k = \bs{q}_k \min \left\{\sqrt{\frac{P}{\sum_{j=1}^K\|\bs{q}_j\|^2}}, 1 \right\}.
\end{equation}

\section{Results}\label{sec:results}
We consider the setup where $T_\tdl<M$ such that the BS only has incomplete CSI. We additionally assume that linear MMSE channel estimation is performed and we initialize all iterative approaches with the zero-forcing precoder $\bs{P}^\text{zf}=\beta\hat{\bs{H}}(\hat{\bs{H}}^\HH \hat{\bs{H}})^{-1}$, with the normalization factor $\beta$ used to satisfy the power constraint.
The results are averaged over 300 channel realizations, where each realization is generated according to $\bs{h}_k\sim\mathcal{N}_\mathbb{C}(\bs{0},\bs{C}_k)$. Here, the channel covariance matrix $\bs{C}_k$ corresponds to a Gaussian mixture model (GMM) \cite{GMM} component. The latter is obtained by fitting a GMM to the collected observations of channels generated with the QUAsi Deterministic RadIo channel GenerAtor (QuaDRiGa) \cite{quadriga}.
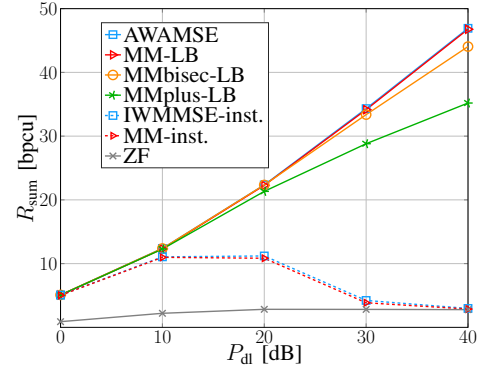
\begin{figure}[H]
	\centering
	\scalebox{0.35}{
%
%
\definecolor{mycolor1}{rgb}{0.07451,0.62353,1.00000}%
\definecolor{mycolor2}{RGB}{255, 140, 0}%
\begin{tikzpicture}

\begin{axis}[%
width=6.028in,
height=4.754in,
at={(1.011in,0.642in)},
scale only axis,
xmin=0,
xmax=40,
ymin=0,
ymax=50,
xlabel style={font=\color{white!15!black},font=\Huge},
xlabel={$P_{\text{dl}}\text{ [dB]}$},
xtick={0,10,20,30,40},
ytick={10,20,30,40,50},
xticklabel style={font=\color{white!15!black},font=\huge},
yticklabel style={font=\color{white!15!black},font=\huge},
axis background/.style={fill=white},
ylabel style={font=\color{white!15!black}, font=\Huge},
ylabel={$R_\text{sum}$ [bpcu]},
xmajorgrids,
ymajorgrids,
legend style={at={(0.1,0.5)}, anchor=south west, legend cell align=left, align=left, draw=white!15!black, font=\Huge}
]

\addplot [color=mycolor1, line width=1.5pt, mark=square, mark options={solid},  mark size=4pt]
  table[row sep=crcr]{%
0	5.08212450638163\\
10	12.3817429098011\\
20	22.3462922077936\\
30	34.3186745156315\\
40	46.9162185093815\\
};
\addlegendentry{AWAMSE}

\addplot [color=red, line width=1.5pt, mark=triangle, mark options={solid,rotate=270},  mark size=5pt]
  table[row sep=crcr]{%
0	5.08444540594126\\
10	12.3060948460459\\
20	22.2770319860073\\
30	34.1526716716698\\
40	46.7694579635684\\
};
\addlegendentry{MM-LB}

\addplot [color=mycolor2, line width=1.5pt, mark=o, mark options={solid,rotate=270},  mark size=5pt]
 table[row sep=crcr]{%
0	5.09025099880932\\
10	12.3777578739676\\
20	22.3411467900474\\
30	33.3717441475338\\
40	44.048896859679\\
};
\addlegendentry{MMbisec-LB}

\addplot[color=green!70!black, line width=1.5pt, mark=star, mark options={solid,rotate=270},  mark size=5pt]
  table[row sep=crcr]{%
0	5.07293685765644\\
10	12.2877853511305\\
20	21.3324702438286\\
30	28.808764411534\\
40	35.1935737698673\\
};
\addlegendentry{MMplus-LB}

\addplot [color=mycolor1, dashed, line width=1.5pt, mark=square, mark options={solid},  mark size=4pt]
  table[row sep=crcr]{%
0	5.06103390449752\\
10	11.0600047503961\\
20	11.2029116493032\\
30	4.22274656821716\\
40	2.96828753124235\\
};
\addlegendentry{IWMMSE-inst.}

\addplot [color=red, dashed, line width=1.5pt, mark=triangle, mark options={solid,rotate=270}, mark size=4pt]
   table[row sep=crcr]{%
0	5.02522395501566\\
10	10.9837953181566\\
20	10.8418920349272\\
30	3.85787316591722\\
40	2.87317477407912\\
};
\addlegendentry{MM-inst.}

\addplot [color=gray, mark=x,mark options={solid},  mark size=5pt, line width=1.5pt]
 table[row sep=crcr]{%
0	0.89189934944006\\
10	2.20596505832931\\
20	2.82920769400413\\
30	2.81659010889233\\
40	2.78227111611381\\
};
\addlegendentry{ZF}
\end{axis}

\end{tikzpicture}%
}
	\caption{Achievable SR versus $P_\tdl$ for $M=32$ antennas, $K=8$ users, $T_\tdl=4$ pilots}
	\label{fig:SR_Tdl=4}
\end{figure}
\par In Fig.~\ref{fig:SR_Tdl=4}, we show the performance of different precoding approaches for a scenario with $M=32$~antennas, $K=8$ users and $T_\tdl=4$ pilots. Obviously, the zero-forcing precoder, as well as the IWMMSE-inst and the MM-inst approaches, which treat the channel estimate as if it were the actual channel, perform poorly, especially as the transmit power increases. We further notice that the proposed MM-LB approach and the AWAMSE method from \cite{VTC24}. Interestingly, the MMplus-LB approach fails to achieve a competitive performance, notably in the high-power region. This might be due to the fact that this approach is based on optimizing a lower bound on the already established lower bound on the sum rate, leading to the performance gap compared to the other approaches, which directly operate on the initial lower bound. Furthermore, the MMbisec-LB approach shows a performance degradation in the high-power region due to the need for a line search of the Lagrangian multiplier. Here, bisection is performed in each precoder update step \cite{bisec}.

\begin{figure}[!ht]
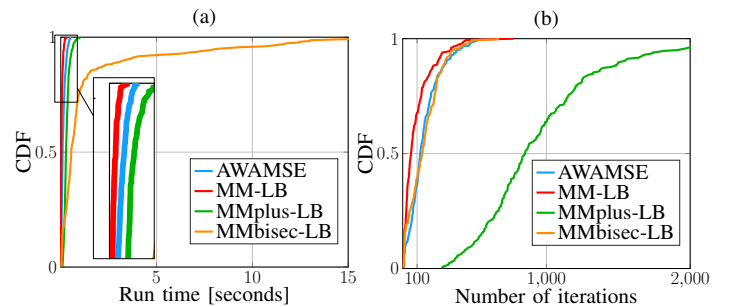

\begin{minipage}{0.48\columnwidth}
    \centering
\scalebox{0.4}{\input{Figs/CDF_Runtime_Pdl=30dB_M=32_K=8_Tdl=4}}
\end{minipage}    
\begin{minipage}{0.1\columnwidth}
\end{minipage}
\begin{minipage}{0.48\columnwidth}
    \centering
\scalebox{0.4}{\input{Figs/CDF_Niter_Pdl=30dB_M=32_K=8_Tdl=4}}
\end{minipage}
\caption{CDFs of the run time (a) and number of iterations (b) for $M=32$ antennas, $K=8$ users, $T_\tdl=4$ pilots and $P_\tdl=30$~dB}
\label{fig:CDF_Tdl=4}
\end{figure}
We now consider the performance in terms of the computational efficiency in Fig.~\ref{fig:CDF_Tdl=4}, where we plot the CDFs of the run time and the number of iterations needed by the different iterative approaches until convergence is reached. The results are obtained for the same setup as before at $P_\tdl=30~$dB. It can be easily seen from Fig.~\ref{fig:CDF_Tdl=4}(a) that the MMbisec-LB approach, where a line search is performed in every iteration, is the most computationally demanding compared to the other approaches with closed-form updates. The CDF of the iteration number in Fig.~\ref{fig:CDF_Tdl=4}(b) shows that the MMplus-LB method requires a high number of iterations to converge, but the run time per iteration is still small such that the overall run time is in the order of the one required for the AWAMSE and MM-LB methods.
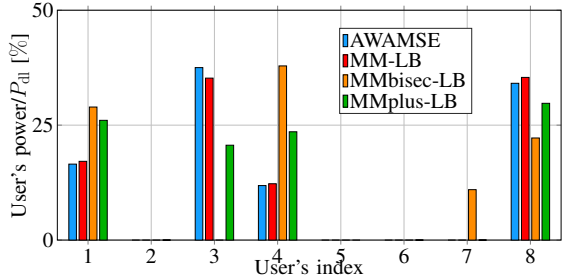
\begin{figure}[H]
	\centering
	\scalebox{0.4}{\definecolor{mycolor2}{RGB}{255, 140, 0}%
\definecolor{mycolor3}{rgb}{0.92900,0.69400,0.12500}%
\definecolor{mycolor1}{rgb}{0.07451,0.62353,1.00000}
\begin{tikzpicture}

\begin{axis}[%
width=6.521in,
height=3in,
at={(0.758in,0.481in)},
scale only axis,
ybar=2.5pt,
bar width=.25cm,
xmin=0.514285714285714,
xmax=8.48571428571429,
xtick={1,2,3,4,5,6,7,8},
yticklabel style={font=\huge},
xticklabel style={font=\huge},
xticklabels={1,2,3,4,5,6,7,8},
ymin=0,
ymax=.5,
ytick={0,.25,.5},
yticklabels={0, 25,50},
ylabel style={font=\color{white!15!black}, font=\huge},
ylabel={User's power/$P_\tdl$ [$\%$]},
xlabel style={font=\color{white!15!black}, font=\huge},
xlabel={User's index},
axis background/.style={fill=white},
xmajorgrids,
ymajorgrids,
legend style={at={(0.56,0.55)}, anchor=south west, legend cell align=left, align=left, draw=white!15!black, font=\huge},
legend image code/.code={
        \draw [#1] (0cm,-0.2cm) rectangle (0.15cm,0.2cm); }
]

\addplot[black,fill=mycolor1, draw=black] coordinates {
(1,	0.1652349803718750) (2,1.796091669102360e-35) (3,0.375159899748188) (4,0.118558609607218)(5,4.991269910711859e-48) (6,8.622646233996124e-40) (7,1.099018609908867e-19) (8,0.341046510272719)
  };

\addplot[black,fill=red, draw=black] coordinates {
(1,0.171291230264419) (2,6.300648584879281e-34) (3, 0.352300266072614) (4,0.122550361132485) (5,2.226160507932428e-48) (6,3.514476142727706e-40) (7,7.461172171164393e-21) (8,0.353858142530481)
  };

\addplot[black,fill=mycolor2, draw=black] coordinates {
(1,0.289407997111003) (2,1.871377701727640e-22) (3,1.815805721113553e-17) (4,0.378853897750142)(5,6.452546793056266e-44) (6,6.469122207186858e-34) (7,0.109662628746658) (8,0.222075476392197)
  };

 \addplot[black,fill=green!70!black, draw=black] coordinates {
(1,0.260489702540119) (2,2.002294207251739e-05) (3,0.206361332768485) (4,0.235571311343201)(5,7.687978388440984e-07) (6,4.140078989126962e-06) (7,4.474355271335984e-05) (8,0.297507977976581)
  };
  
\legend{AWAMSE, MM-LB, MMbisec-LB, MMplus-LB}

\end{axis}

\end{tikzpicture}%
}
	\caption{Power allocation for $P_\tdl=40$~dB, $M=32$ antennas, $K=8$ users, $T_\tdl=4$ pilots}
	\label{fig:powers}
\end{figure}
The different behavior in the high power regime is also confirmed by the power allocation results shown in Fig.~\ref{fig:powers}. Firstly, we need to recall the finding from \cite[Subsection~II-B]{ZF_LS} that the number of active users is upper bounded by $T_\tdl$. Obviously, for all methods, there are $K^\text{active}=T_\tdl=4$ active users. However, we can see that MMbisec-LB exhibits an allocation pattern different from those of the other approaches, leading to the performance gap compared to AWAMSE and MM-LB. The latter two show a similar power allocation among the active users. MMplus-LB serves the same users as AWAMSE and MM-LB but distributes the transmit power among them in a considerably different manner.

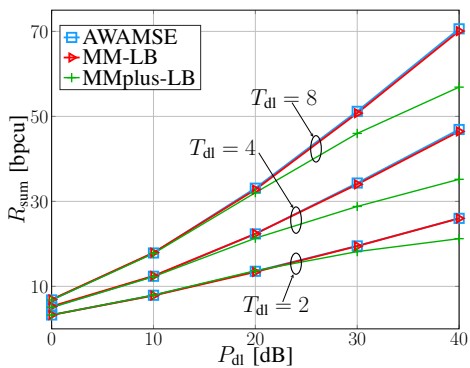
\begin{figure}[!ht]
	\centering
	\scalebox{0.35}{
%
%
\definecolor{mycolor1}{rgb}{0.07451,0.62353,1.00000}
\begin{tikzpicture}

\begin{axis}[%
width=6.028in,
height=4.754in,
at={(1.011in,0.642in)},
scale only axis,
xmin=0,
xmax=40,
ymin=0,
ymax=75,
xlabel style={font=\color{white!15!black},font=\Huge},
xlabel={$P_{\text{dl}}\text{ [dB]}$},
xtick={0,10,20,30,40},
ytick={10,30,50,70},
xticklabel style={font=\color{white!15!black},font=\huge},
yticklabel style={font=\color{white!15!black},font=\huge},
axis background/.style={fill=white},
ylabel style={font=\color{white!15!black}, font=\Huge},
ylabel={$R_\text{sum}$ [bpcu]},
xmajorgrids,
ymajorgrids,
legend style={at={(0.02,0.75)}, anchor=south west, legend cell align=left, align=left, draw=white!15!black, font=\Huge}
]




\addplot [color=mycolor1, line width=2pt, mark=square, mark options={solid},  mark size=5pt]
  table[row sep=crcr]{%
0	6.81050926038937\\
10	17.8582098464894\\
20	33.106273397664\\
30	51.1899713928343\\
40	70.6049654041712\\
};
\addlegendentry{AWAMSE}

\addplot [color=red, line width=2pt, mark=triangle, mark options={solid,rotate=270},  mark size=5pt]
  table[row sep=crcr]{%
0	6.92330767484351\\
10	17.9339010566865\\
20	32.7626545116533\\
30	50.7568494630698\\
40	70.079643320937\\
};
\addlegendentry{MM-LB}

\addplot [color=green!70!black, line width=1.5pt, mark=+, mark options={solid},  mark size=5pt]
 table[row sep=crcr]{%
0	6.79606836134082\\
10	17.7358460657705\\
20	32.053074064949\\
30	45.9883475970539\\
40	56.8823911174671\\
};
\addlegendentry{MMplus-LB}

\addplot [color=mycolor1, line width=2pt, mark=square, mark options={solid},  mark size=5pt]
  table[row sep=crcr]{%
0	5.08212450638163\\
10	12.3817429098011\\
20	22.3462922077936\\
30	34.3186745156315\\
40	46.9162185093815\\
};

\addplot [color=red, line width=2pt, mark=triangle, mark options={solid,rotate=270},  mark size=5pt]
  table[row sep=crcr]{%
0	5.29826010649866\\
10	12.4686740812024\\
20	22.3827145789822\\
30	34.0088677350462\\
40	46.4698538788978\\
};

\addplot [color=green!70!black, line width=1.5pt, mark=+, mark options={solid},  mark size=5pt]
  table[row sep=crcr]{%
0	5.07293685765644\\
10	12.2877853511305\\
20	21.3324702438286\\
30	28.808764411534\\
40	35.1935737698673\\
};


\addplot [color=mycolor1, line width=2pt, mark=square, mark options={solid},  mark size=5pt]
  table[row sep=crcr]{%
0	3.26989999134003\\
10	7.9055685734763\\
20	13.549087882846\\
30	19.5216077002265\\
40	26.0075132671108\\
};

\addplot [color=red, line width=2pt, mark=triangle, mark options={solid,rotate=270},  mark size=5pt]
  table[row sep=crcr]{%
0	3.27524537404456\\
10	7.84332328128343\\
20	13.399009528521\\
30	19.4727236017599\\
40	25.9998161422032\\
};

\addplot [color=green!70!black, line width=1.5pt, mark=+, mark options={solid},  mark size=5pt]
  table[row sep=crcr]{%
0	3.25569586922858\\
10	7.95336747651046\\
20	13.6906191735266\\
30	18.1809342382448\\
40	21.2269230017941\\
};

\end{axis}

\draw [black, very thick] (12.5, 8.45) ellipse [x radius=0.2, y radius=0.5];
\draw [black, very thick] (11.75,5.75) ellipse [x radius=0.2, y radius=0.5];
\draw [black, very thick] (11.75,4.1) ellipse [x radius=0.2, y radius=.4];

\draw [-stealth, thick](11.6,10.2) -- (12.4,8.9);
 \node at (11.3,10.35) {\Huge $T_\text{dl}=8$};
\draw [-stealth, thick](10.35,8.3) -- (11.65,6.2);
\node at (9,8.45) {\Huge $T_\text{dl}=4$};
\draw [-stealth, thick](11.25,2.65) -- (11.65,3.7);
\node at (11,2.5) {\Huge $T_\text{dl}=2$};

\end{tikzpicture}%
}
	\caption{Achievable SR versus $P_\tdl$ for $M=32$ antennas, $K=8$ users and different pilot numbers}
	\label{fig:SR_Tdls}
\end{figure}
In Fig.~\ref{fig:SR_Tdls}, we now show the results for different numbers of pilots, namely $T_\tdl=2,4,8$. One can easily see that MM-LB and AWAMSE perform almost in the same manner for all three cases. However, the MMplus-LB approach does not deliver comparable performance in the high-power regime even when the number of pilots is increased.




\section{Conclusion}
This paper presented a robust precoding in FDD MISO systems with imperfect CSI. The proposed MM-LB method leverages minorization-maximization techniques to construct and optimize a lower bound on the achievable sum rate.  
We demonstrated the competitive sum rate performance compared to state-of-the-art methods while maintaining low computational complexity due to the closed-form update of the precoder. This makes it a promising candidate for practical implementation in large-scale MISO systems with limited CSI.
	

	\bibliographystyle{unsrt} 


\end{document}